\documentclass[final,3p,times,twocolumn]{elsarticle}

\usepackage{graphicx, epsfig, amssymb}

\journal{Computer Physics Communications}

\begin{document}

\begin{frontmatter}



\title{Can we predict the failure point of a loaded composite material ?}


\author{Srutarshi Pradhan}

\address{Formation Physics Department, SINTEF Petroleum Research, NO-7465 Trondheim, Norway}
\ead{srutarshi.pradhan@sintef.no}

\begin{abstract}
As a model of composite material, the fiber bundle model has been 
chosen -where a bundle of 
fibers is subjected to external load and fibers have distributed thresholds.
For different loading conditions, such a system shows few precursors which 
indicate that the complete failure is imminent.
When external load is increased quasi-statically - \textit{bursts}
 (number of failing fibers) of different sizes are produced. The burst 
statistics shows a robust 
crossover behavior near the failure point, around which the average burst size 
seems to diverge. If the load is increased by discrete steps, 
susceptibility and relaxation time diverge as failure point is approached.  
When the bundle is overloaded (external load is more than critical load) the 
rate of breaking shows a minimum at half way to the collapse point. The 
pattern and statistics of energy emission bursts show characteristic 
difference for below-critical and over-critical load levels.
\end{abstract}

\end{frontmatter}


\section{Introduction}

Prediction of the failure point is a major challenge in various scenarios of 
fracture and breakdown \cite{BKC-book,Herrmann-book,PB-book} --ranging from 
fracturing in nano-materials  to onset of  earthquakes. 
A material body can tolerate certain level of load or force on it and beyond 
that level it collapses. If the load is increased continuously - when does 
the collapse point come ? Is there any precursor which signals that the 
complete failure is imminent ?
    
Fiber bundle model (FBM) has been proved 
\cite{Peirce, Daniels,Phoenix,pradhanRMP} to capture the essentials of 
failures in composite materials. FBM contains a large number of fibers with 
statistically distributed strength thresholds. It has simple geometry and 
clear-cut rules for how stress caused by a failed fibers is redistributed on 
intact fibers.  Most importantly, this model can be solved analytically to an 
extent (For reviews, see \cite{pradhanRMP}) that is not possible for any other 
model of material-failure. The statistical properties of FBM is 
well studied \cite{HH,PHH05,HHP}, and the failure 
dynamics at a constant load has been formulated \cite{PBC}  through recursion
 relations which in turn explore the phase transition and associated critical 
behavior in this model.

\begin{figure}
\includegraphics[width=5.5cm,height=4.5cm]{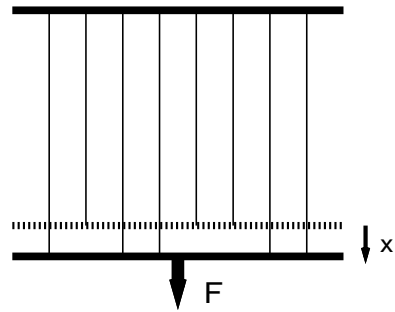}
\caption {The fiber bundle model of composite materials. }
\label{FBM-model}
\end{figure}

In this article we discuss how we can predict the failure point of a loaded 
FBM from the available precursors. 
The term \textit{precursor} usually means \cite{PC06} some prior indications of 
an upcoming incident and in current context such incident is the complete 
failure (collapse) of a fiber bundle under external load.  
We consider equal-load-sharing model, in 
which the load previously carried
by a failed fiber is shared equally by all the remaining intact fibers
\cite{pradhanRMP}. The
bundle consisting of $N$ elastic fibers, clamped
at both ends (Fig. \ref{FBM-model}). All the fibers obey Hooke's law with force 
constant set to unity for simplicity.
Each fiber $i$ is associated with a breakdown threshold $x_{i}$
for its elongation. When the length exceeds $x_{i}$ the fiber breaks
immediately, and does not contribute to the strength of the bundle
thereafter. The individual thresholds $x_{i}$ are assumed to be independent
random variables with the same cumulative distribution function $P(x)$
and a corresponding density function $p(x)$: 
\begin{equation}
{\rm Prob}(x_{i}<x)=P(x)=\int_{0}^{x}p(y)\; dy.
\end{equation}

We analyse three different loading cases: quasi-static loading, 
load increment by equal steps and overloaded situation. For prediction 
purpose it is important that precursors can be seen in a single system. 
Therefore, throughout this article we will present and discuss results that 
can be seen in a single bundle containing large number of fibers. For 
simplicity, we consider the uniform distribution of fiber thresholds:
 $P(x)=x$ for $0\leq x\leq 1$.

\section{ Quasi-static loading}
The \emph{quasi-static} loading is a  strain controlled method, 
where at each step the whole bundle is stretched till the weakest fiber 
(among the intact ones) fails. 
At an elongation $x$ per surviving fiber the total force on the bundle
is $x$ times the number of \textit{intact} fibers. The expected or average
force at this stage is therefore \begin{equation}
F(x)=N\, x\,(1-P(x)).\label{load}\end{equation}
The maximum $F_{c}$ of
$F(x)$ corresponds to the value $x_{c}$ for which $dF/dx$ vanishes.
Thus \begin{equation}
1-P(x_{c})-x_{c}p(x_{c})=0;\end{equation}
where $x_{c}$ is the critical elongation value  above which the bundle 
collapses. 

\subsection{Burst or avalanche of failing fibers}
When a fiber fails, the stress on the intact fibers increases.
This may in turn trigger further fiber failures, which can produce
 bursts (avalanches) that either lead to a stable situation or to breakdown
of the whole bundle. A burst is usually defined as the amount
or number ($\Delta$) of simultaneous fiber failure during loading.
It was shown in  \cite{HH}
that the average number of burst events
of size $\Delta$  follows a power law of the form
\begin{equation}
\label{eq9}
D(\Delta)/N = C \Delta^{-\xi}
\end{equation}
in the limit $N \to \infty$.  Here, $\textstyle \xi={5\over 2}$
is the {\it universal\/} burst exponent and $C$ is a constant.  
The value of $\xi$
is, under  mild assumptions, independent of the threshold distribution
$P(x)$: the probability density needs to have a quadratic maximum somewhere 
within the range of threshold values.

\subsection{ The crossover behavior}
\begin{figure}[tb]
\includegraphics[width=5.5cm,height=4.5cm]{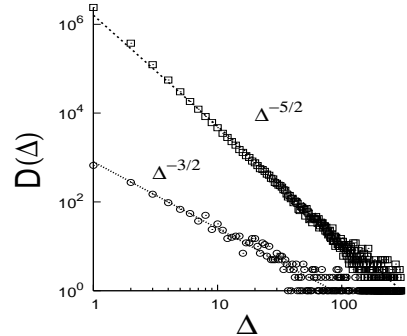}
\caption{The burst distributions: 
all bursts (squares) and bursts within an interval $0.9x_{c}$ and  $x_c$  
(circles). The figure is based on a single bundle containing $N=10^{7}$ fibers 
with uniformly distributed fiber thresholds within $0$ and $1$.}
\label{single-dist}
\end{figure}

When all the bursts are recorded for the entire failure process, the burst 
distribution  follows the asymptotic
power law $D(\Delta) \propto\Delta^{-5/2}$. If we just sample bursts that
occur near the breakdown point, a different power law is seen --which can 
be explained analytically \cite{PHH05,PHH06}.
Fig. \ref{single-dist} compares the complete burst distribution with what we 
get when we sample merely bursts  in the threshold interval
$(0.9x_{c},x_{c})$.  

This observation may be of practical importance, as it gives a
criterion for the imminence of complete failure \cite{PHH05}.
The bursts or avalanches can be
recorded from outside -without disturbing the ongoing failure process.
Therefore, any signature in burst statistics that can warn of imminent system
failure would be very useful in the sense of wide scope of applicability.
It is enticing to note the recent observation  \cite{kawamura06} of a crossover
behavior in the magnitude distribution of earthquakes before the largest
earthquake appears. A similar crossover behavior is also 
seen \cite{PHH06,HHP} in the burst distribution and energy distribution of
the fuse model which is a standard model for studying fracture and
breakdown phenomena in disordered systems. 
Most important is that this
crossover signal does not hinge on observing rare events and is seen  in
a single system (see Fig. \ref{single-dist}). Therefore, such crossover 
signature has a strong potential to be used as a detection tool.

\subsection{Variation of average burst size}
\begin{figure}[t]
\includegraphics[width=5.5cm,height=4.5cm]{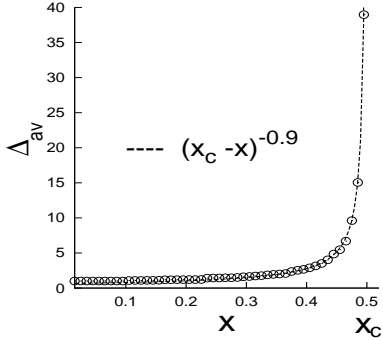}
\caption{Average burst size $\Delta_{av}$ vs. elongation $x$  
for the same fiber bundle as in Fig. \ref{single-dist}.} 
\label{avdel2}
\end{figure}

\begin{figure}[t]
\includegraphics[width=5.5cm,height=4.5cm]{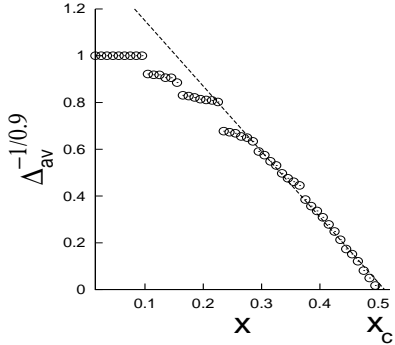}
\caption{Inverse of average burst size is plotted against  $x$  for the same 
data set as in Fig. \ref{avdel2}. A straight line can be fitted near $x_{c}$ 
from which  one can predict the failure point.}  
\label{avdel1}
\end{figure}

We have seen that if external load is increased quasi-statically on a bundle 
of large number of fibers, bursts of different sizes occur during the whole
breaking process till complete failure. One can ask - what is the average 
burst size at a particular elongation ($x$) value? The average burst size is 
indeed a very relevant
quantity that can be measured easily during the failure process. It seems 
(Fig. \ref{avdel2}) that average burst size( $\Delta _{av}$) around some 
elongation value $x$ goes as 
\begin {equation} 
\Delta_{av}(x) \simeq (x_{c}-x)^{0.9}.
\end{equation} 
This means if we plot $\Delta_{av} ^{1/0.9}$ vs. $x$, we should get a 
straight line which touches the $X$ axis at $x=x_{c}$. Even in a single system 
we can see this signature (Fig. \ref{avdel1}).    

\section{Load increment by equal steps}
In the force-controlled case, load can be increased on a bundle  
by equal amount at each loading step. Here, the failure dynamics of 
the bundle can be represented by a recursion relation: 
 Let $N_t$ be the number of fibers  that survive after  step $t$, where $t$
indicates the number of stress redistribution steps.
We call  $\sigma=F/N$, the applied stress and $U_{t}=N_t/N$, the
 surviving fraction of total fibers.
Then the effective stress  after $t$  step becomes
$x_{t}=\frac{\sigma}{U_{t}}$ and after $t+1$  steps the
surviving fraction of total fibers will be $U_{t+1}=1-P(x_{t})$. Therefore we 
can construct the following  recursion relation \cite{PBC}
 \begin{equation}
U_{t+1}=1-P(\sigma/U_{t});U_{0}=1.\label{rec-U}\end{equation}
At equilibrium  $U_{t+1}=U_{t}\equiv U^{*}$.
For uniform fiber strength distribution,
the cumulative distribution becomes
$P(\sigma/U_{t})=\sigma /U_{t}$.
Therefore the  recursion relation becomes
\begin{equation}
U_{t+1}=1-\frac{\sigma}{U_{t}}.\label{rec-U-uniform}
\label{Rec-basic}
\end{equation}
At the fixed point the above relation takes a quadratic form
$U^{*^{2}}-U^{*}+\sigma=0,$
with the solutions
\begin{equation}
U^{*}(\sigma)=\frac{1}{2}\pm(\sigma_{c}-\sigma)^{1/2}; \sigma_c =\frac{1}{4}.
\label{sol-U-uniform}
\end{equation}
One can define the breakdown susceptibility $\chi$,
as the change of $U^{*}(\sigma)$ due to an infinitesimal
increment of the applied stress $\sigma$: \begin{equation}
\chi=\left|\frac{dU^{*}(\sigma)}{d\sigma}\right|=\frac{1}{2}(\sigma_{c}-\sigma)^{-\beta};\beta=\frac{1}{2}.\label{sus-uniform}\end{equation}
To study the failure dynamics around the critical point 
($\sigma\rightarrow\sigma_{c}$), the recursion relation 
(Eq. \ref{rec-U-uniform}) can be replaced
by a differential equation \begin{equation}
-\frac{dU}{dt}=\frac{U^{2}-U+\sigma}{U}.\label{diff-uniform}\end{equation}
Close to the fixed point one gets \cite{PBC} 
\begin{equation}
U_{t}(\sigma)-U^{*}(\sigma)\approx\exp(-t/\tau),\end{equation}
 where the relaxation time 
$\tau=\frac{1}{2}\left[\frac{1}{2}(\sigma_{c}-\sigma)^{-1/2}+1\right]$.
Therefore, near the critical point $\tau$ diverges: 
\begin{equation}
\tau\propto(\sigma_{c}-\sigma)^{-\theta};\theta=\frac{1}{2}.\label{tau-uniform}\end{equation}

\begin{figure}[t]
\includegraphics[width=5.5cm,height=4.5cm]{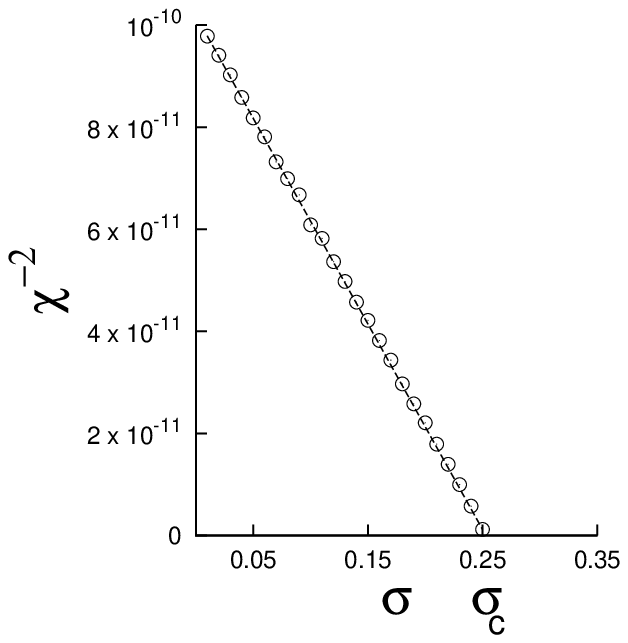}
\includegraphics[width=5.5cm,height=4.5cm]{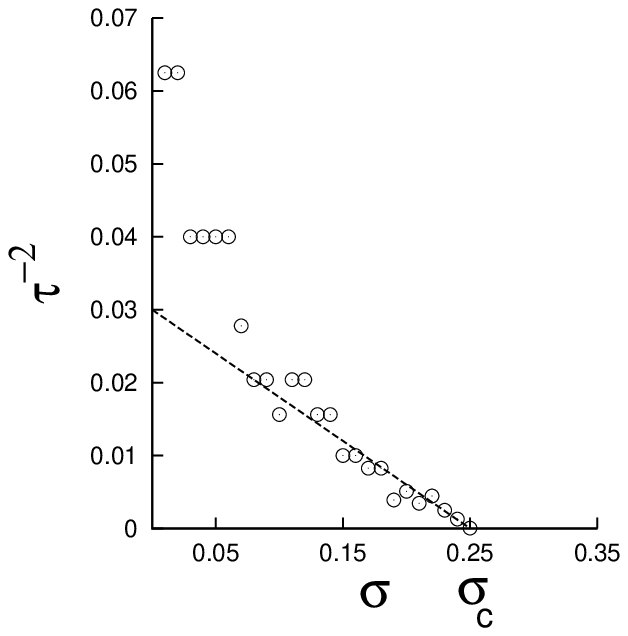}
\caption{ Variation of $\chi^{-2}$ and $\tau^{-2}$
with applied stress 
for the same bundle as in Fig.\ref{single-dist}. 
The dotted straight lines are the best linear fits near the critical point.}
\label{fig:xi-tau}
\end{figure}

Since the susceptibility ($\chi$) and the relaxation time ($\tau$)
follow power laws (exponent $= -1/2$) with external stress and both  diverge at
the critical stress, therefore, if we plot $\chi^{-2}$ and $\tau^{-2}$
with external stress, we expect a linear fit near critical point and
the straight lines should touch $X$ axis at the critical stress.
We indeed found similar behavior (Fig. \ref{fig:xi-tau}) in simulation
experiments for a single sample.

\section{ The over-loaded situation}
\begin{figure}[t]
\includegraphics[width=5.5cm,height=4.5cm]{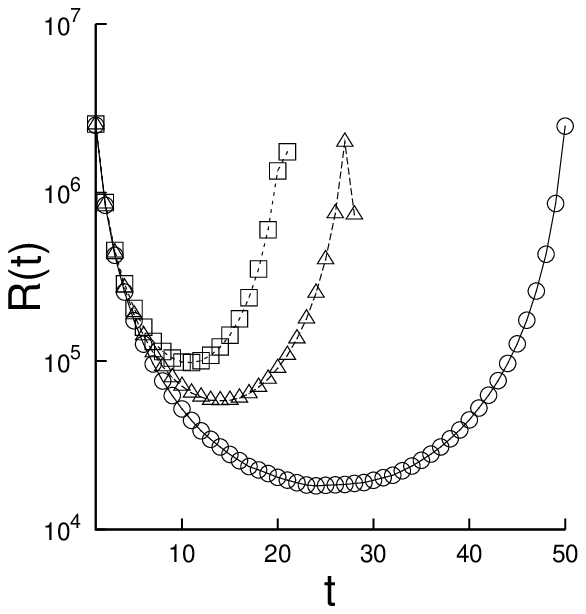}
\includegraphics[width=5.5cm,height=4.5cm]{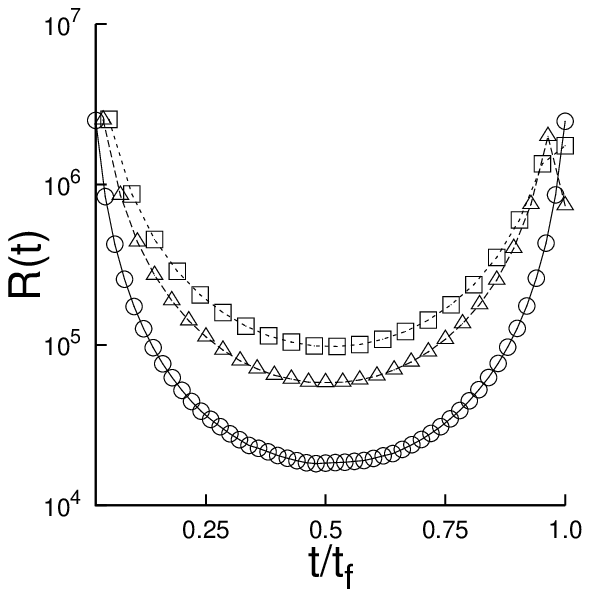}
\caption { Breaking rate $R(t)$ vs. step $t$ (upper plot) 
and vs.\ the rescaled step variable $t/t_f$ (lower plot) 
for the same fiber bundle as in Fig.\ref{single-dist}. 
Different symbols are used for different excess stress levels 
$\sigma -\sigma _c$: 0.001  (circles), 0.003 (triangles) and 0.005 (squares). }
\label{breaking-rate-fig}
\end{figure}

What happens if the initial applied load  $F=N\sigma$  is larger than the 
critical load of the bundle ? 
The stepwise failure process continues and the bundle collapses at some 
step $t_f$. 
If we consider the uniform distribution of fiber thresholds,
and assume that the load is slightly above the 
critical value: $\sigma = \sigma_c +\epsilon,$
with $\epsilon>0$,
 we can rewrite  the recursion relation (Eq. \ref{Rec-basic}) as
\begin{equation}
U_{t+1} = 1 - ({\textstyle \sigma_c}+\epsilon)\frac{1}{U_t}.
\label{unin}\end{equation}
The solution \cite{PH09} of the above recursion shows  that 
there is a relation between the minimum of the breaking rate 
$R(t)=dU_t/dt$ (treating $t$ as continuous variable) and the final step  $t_f$: \begin{equation}
t_f = 2 t_0.
\end{equation}
At the minimum of the breaking rate, the bundle is just  halfway to its 
complete collapse.
Simulations on a single sample show that the breaking rate has a  
minimum at some value $t_0(\epsilon)$, and that for varying $\epsilon$  the 
minima all occur at a value close to $\frac{1}{2}$ when plotted as function 
of the scaled variable $t/t_f$ (Fig.\ \ref{breaking-rate-fig}). 

\section{ Energy emission bursts of loaded FBM}
\begin{figure}[t]
\includegraphics[width=5cm,height=3.8cm]{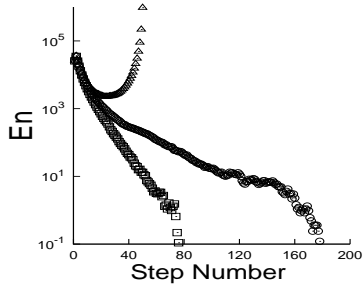}
\caption {Energy emission  $En$ vs. step number 
in the same bundle as in Fig.\ref{single-dist}. 
Different symbols indicate different stress levels: at critical stress $\sigma_c$ (circles); stress $\sigma_c +0.001$ (triangles) and stress $\sigma_c -0.001$ (squares).}
\label{En-pattern}
\end{figure}

\begin{figure}[t]
\includegraphics[width=5cm,height=3.8cm]{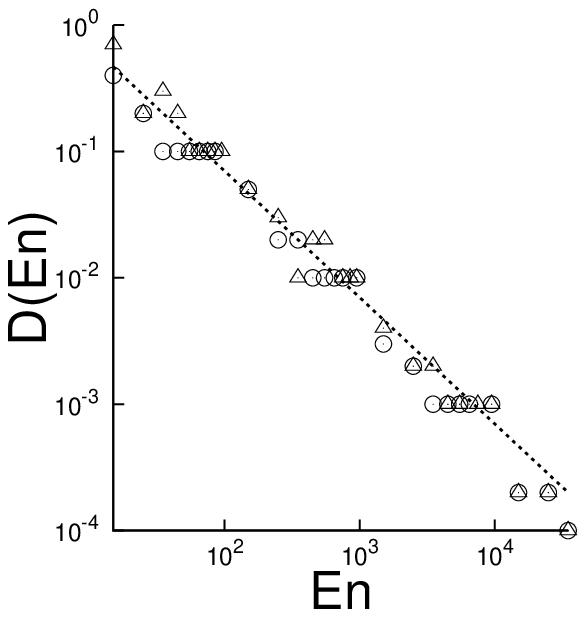}
\caption {Distribution of energy emissions  $En$ in the same fiber bundle as in 
Fig \ref{single-dist} for stresses below 
critical value: $\sigma_c -0.001$ (circles), $\sigma_c -0.002$ (triangles). 
The straight line has a slope $-1.0$. }
\label{En-below}
\end{figure}

\begin{figure}[h]
\includegraphics[width=5cm,height=3.5cm]{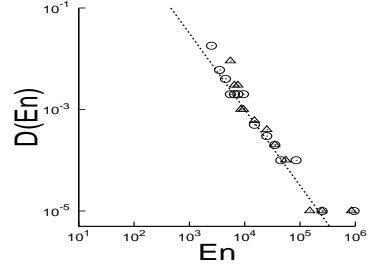}
\caption {Distribution of energy emissions  $En$ in the same bundle as in 
Fig. \ref{single-dist} for stresses above the 
critical value: $\sigma_c +0.001$ (circles), $\sigma_c +0.002$ (triangles). 
The straight line has a slope $-1.5$.}
\label{En-over}
\end{figure}

It is well known that during fracturing process in composite materials, 
energy releases in the form of acoustic emissions 
\cite{Petri, Ciliberto97} and most of the cases these acoustic bursts 
follow power laws.  
Very recently the statistics of energy emission bursts has been studied 
\cite{PH08} in FBM -both analytically and through numerical simulations. 
As the fibers obey Hooks law up to the failure, when a fiber fails at an 
elongation $x$, elastic energy of amount $\frac{1}{2} K x^2$ will be released, 
where $K$ is the force constant.
Therefore for a burst of size $\Delta$ the corresponding energy release $En$ 
can be calculated as
\begin{equation}
En = \frac{1}{2} K \sum^{\Delta+min} _{i=min}{x_i^2},  
\end{equation}  
where   $x_i$ is the strength of failing fiber. Now at a constant applied load,
if we record $En$ at each step of load redistribution, it shows different 
pattern (Fig. \ref{En-pattern}) depending on the stress level --at critical, 
over-critical or below-critical. If we record such energy emission bursts 
separately for below-critical and over-critical levels, the corresponding 
distributions exhibit convincing 
difference (Figs. \ref{En-below},\ref{En-over}). For stresses below-critical
level, there are many many  small energy bursts which are absent in the 
later case. The exponent of the energy burst 
distributions are different for these two situations: $-1$ and 
$-1.5$ respectively.    

\section {Summary and discussions}
In summary,  the fiber bundle model of composite material shows some precursors
which can help to predict the failure point of the system under different 
loading situations.
When external load is increased quasi-statically, near the failure point the 
exponent of burst distribution shows a crossover from $-5/2$ to $-3/2$  and
 the average burst size  diverges. While the crossover signature can only warn 
of an imminent failure, 
it is possible to predict critical elongation value ($x_c$) in advance by 
measuring the average burst size. If the load is increased by equal steps, 
susceptibility and relaxation time seem to diverge at failure stress 
($\sigma_c$) following robust power laws - from which one can predict the 
failure stress value without approaching too close to the failure point.
When the bundle is overloaded the 
rate of breaking has a minimum at half way to the collapse point and the
 distributions of energy emission bursts follow different power laws for
 below-critical and over-critical load levels. A very recent study \cite{PH10}
 shows that for overloaded case, the energy 
bursts also attain a minimum value around half way to complete collapse.

\section {Acknowledgement}
The author acknowledges financial support from Research Council of Norway 
(NFR) through project number 199970/S60.  

\bibliographystyle{elsarticle-num}

\end{document}